\documentclass[%
 reprint,
 amsmath,amssymb,
 aps,
 prl,
floatfix
]{revtex4-2}

\usepackage{graphicx}
\usepackage{dcolumn}
\usepackage{bm}
\usepackage{upgreek}
\usepackage{xcolor}

\begin{document}


\title{Temperature gradient-driven motion of magnetic domains\\
in a magnetic metal multilayer by entropic forces}
 
\author{Lin Huang}
\altaffiliation[Present address: ]{School of Material Science, University of Sheffield, Sheffield S10 2TN, United Kingdom.}
\affiliation{School of Physics and Astronomy, University of Leeds, Leeds LS2 9JT, United Kingdom.}

\author{Joseph Barker}
\affiliation{School of Physics and Astronomy, University of Leeds, Leeds LS2 9JT, United Kingdom.}
 
\author{Lekshmi Kailas}
\affiliation{School of Physics and Astronomy, University of Leeds, Leeds LS2 9JT, United Kingdom.}

\author{Soumyarup Hait}
\affiliation{School of Physics and Astronomy, University of Leeds, Leeds LS2 9JT, United Kingdom.}

\author{Simon D.~Connell}
\affiliation{School of Physics and Astronomy, University of Leeds, Leeds LS2 9JT, United Kingdom.}

\author{Gavin Burnell}
\affiliation{School of Physics and Astronomy, University of Leeds, Leeds LS2 9JT, United Kingdom.}

\author{Christopher H.~Marrows}
\email[Email:~]{c.h.marrows@leeds.ac.uk}
\affiliation{School of Physics and Astronomy, University of Leeds, Leeds LS2 9JT, United Kingdom.}
  
\date{\today}

\begin{abstract}
We studied the displacement of magnetic domains under temperature gradients in perpendicularly magnetized Ta/[Pt/Co$_{68}$B$_{32}$/Ir]$_{\times 10}$/Pt multilayer tracks with microfabricated Pt heaters/thermometers by magnetic force microscopy (MFM). Subtracting out the effects of the Oersted field from the heating current reveals the pure temperature gradient-driven motion, which is always towards the heater. The higher the thermal gradient along the track (owing to proximity to the heater or larger heater currents), the greater the observed displacements of the domains, up to a velocity of around 1~nm/s in a temperature gradient of 20~K/$\upmu$m. Quantitative estimates of the strength of different driving mechanisms show that entropic forces dominate over those arising from the spin Seebeck and spin-dependent Seebeck effects. 
\end{abstract}

\maketitle



The field of spin caloritronics~\cite{Bauer2012,Adachi2013,Yu2017,Uchida2021} lies at the intersection of thermoelectric effects and spintronics. Although it is well known that the dynamics of spin textures such as domain walls (DWs)~\cite{Marrows2005,Ralph2008,Thiaville2012} and skyrmions~\cite{Jonietz2010,Jiang2015,Woo2016,Zeissler2020} can be driven by the flow of currents down electrical potential gradients, spin caloritronics opens up the possibility that motion can be driven by temperature gradients, $\nabla T$. 

Three principal mechanisms are proposed for this. First, in the spin Seebeck effect~\cite{Uchida2010,Jaworski2010,Uchida2014} a current of magnons flows down a temperature gradient. 
This can take place in an insulator or metal, and the current of angular momentum can exert a spin transfer torque (STT) on a spin texture~\cite{Kovalev2009,Slonczewski2010,Hinzke2011,Yan2011,Wang2012,Kovalev2012,Yan2015,Akanda2023}. Second, the spin-dependent Seebeck effect takes place in metals, in which a pure spin current of electrons flows down the temperature gradient~\cite{Uchida2008,Yi2020}. Again, this spin current can exert spin-transfer torques on spin textures~\cite{Wang2022}. Third, entropic forces arise due to the gradient of temperature-dependent magnetic parameters such as magnetization $M_\mathrm{sat}$ or exchange stiffness $A$, which determine the energy of the spin texture~\cite{Schlickeiser2014}. Although atomistic simulations of a DW in an Fe monolayer on W using a Landau-Lifshitz-Gilbert approach were interpreted only in terms of magnon current flows~\cite{Chico2014}, micromagnetic simulations of a Permalloy wire using the Landau-Lifschitz-Bloch equation suggest that these entropic effects are likely to be the predominant driving force for DW motion~\cite{Moretti2017}. 

Experimentally, early evidence for thermal STTs included the current dependence of switching fields~\cite{Yu2010} and the increased likelihood of finding a DW in a spot in the sample heated by a laser~\cite{Moehrke2010,Tetienne2014,Ramsay2015}. Analysis of the mechanism at play was limited or of a qualitative nature. DW motion has been directly imaged in a yttrium iron garnet (YIG) film using the magneto-optical Kerr effect, and was interpreted purely in terms of magnon currents~\cite{Jiang2013}. The unidirectional component of DW motion under a current pulse in a (metallic) permalloy wire has also been interpreted in terms of the magnon spin transfer picture~\cite{Torrejon2012}. In all of these experiments, the DW moves towards the hot region. For skyrmions, the picture is less clear. Motion of a skyrmion lattice towards hotter regions has been observed in Cu$_2$OSeO$_3$~\cite{Yu2021}, but towards colder regions in [Pt/CoFeB/Ta]$_{15}$ multilayers~\cite{Wang2020}. This discrepancy motivated the work of Raimondo~\textit{et al.} in the theoretical study of how different temperature scalings of the magnetic free energy terms lead to entropic forces \cite{Raimondo2022}.

In this Letter, we report a study of domain motion in a perpendicularly magnetized Pt/CoB/Ir metal multilayer driven by a temperature gradient. Our principal experimental result is that the domains always move towards the hotter region once a critical value of temperature gradient is exceeded. We observe larger displacements of the domains closer to the heater, where $\nabla T$ is larger, allowing us to map the temperature profile. Estimates of the strength of the different driving forces that lead to domain motion, based on our experimentally measured parameters, indicate that motion driven by entropic forces dominates over spin-transfer torques arising from currents of both magnons and electrons, confirming the prediction of Moretti~\textit{et al.}~\cite{Moretti2017}. 

\begin{figure}
    \includegraphics[width=7cm]{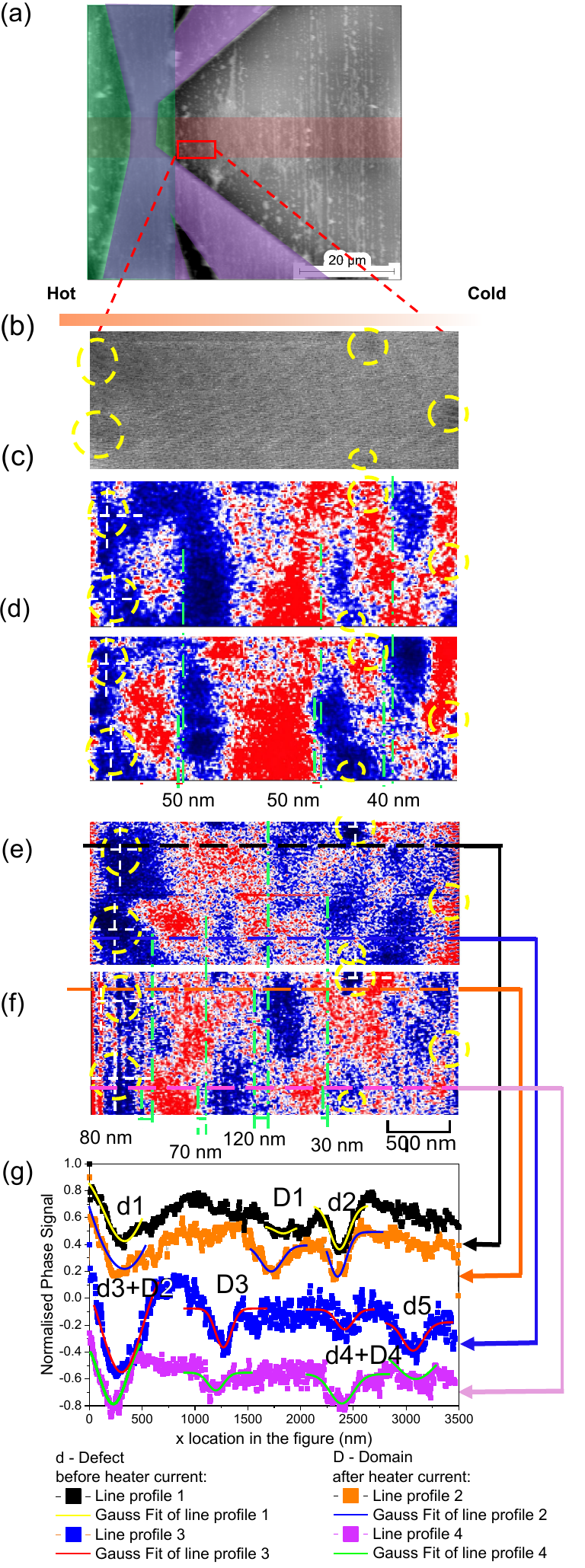}
    \caption{MFM imaging of thermally-induced domain motion. (a) Optical micrograph of magnetic track (red) spanned by electrically isolated (green) Pt heater/thermometer (purple). (b) MFM image when the track is fully saturated in $+700$ Oe, showing defect positions in dashed yellow circles. (c) and (d) are the MFM images in $+30$~Oe before and after a $+30$~mA current was applied to the heater, respectively.  (e) and (f) MFM images in $+30$~Oe before and after a $-30$ mA current was applied to the heater, respectively. Dashed green lines indicate position of the leading edge of the reverse domains. Dashed horizontal lines indicate the positions of the line profiles that are shown in (g), offset for clarity. Gaussian fits to the line profile dips (marked `D' for a domain and `d' for a defect) are indicated by the solid coloured lines.
    \label{fig:30mA}}
\end{figure}

A false color image of our sample is shown in Fig.~\ref{fig:30mA}(a). The 10~$\upmu$m wide magnetic track (red region) was patterned from a Ta(20~\AA)/[Pt(7~\AA)/Co$_{68}$B$_{32}$(8~\AA)/Ir(5~\AA)]$_{\times 10}$/Pt(14~\AA) multilayer that is perpendicularly magnetized. The track was partly covered by 100~nm of insulating SiO$_x$ (green region) to electrically isolate it from a 250~nm thick Pt heater/thermometer wire (purple region). The SiO$_x$ was deposited by RF sputtering, the metals by dc sputtering. All patterning was performed by conventional photolithography. In the Supplementary Material we show a hysteresis loop of a sheet film of equivalent material that confirms perpendicular magnetization~\cite{suppmater}, along with typical domain patterns. 

 
We used magnetic force microscopy (MFM) to observe the magnetic domains and their motion under magnetic fields $H$ applied perpendicular to the sample plane. Fig.~\ref{fig:30mA}(b) shows an MFM image acquired at $H = +700$~Oe, strong enough to magnetically saturate the track. The region of the track imaged here is indicated by the red box in Fig.~\ref{fig:30mA}(a). In this saturated state, ideally the magnetic contrast must be uniform. Therefore we can identify five patches of contrast as defects which are marked by the dashed yellow circles. We used these defects as reference points for measuring the motion of magnetic domains. 

Fig.~\ref{fig:30mA}(c) shows an MFM image after saturation at $+600$~Oe and acquired at $H = +30$~Oe, where regions of blue contrast corresponding to reverse domains are evident. Drift in the MFM imaging of the measured area reduces the number of defects we can resolve, but the same area can still be identified.  Fig.~\ref{fig:30mA}(d) shows the same area after passing a $+30$~mA current, $I$, through the heater for two minutes whilst holding $H$ constant at $+30$~Oe. This generates a temperature gradient down the track as indicated above panel (b). The dashed green lines indicate the motion of the leading edges of the domains towards the heater by a few tens of nanometres. 

The domains were erased and re-nucleated by saturating the sample in $H > 700$~Oe and then decreasing $H$ back to $+30$~Oe. The domains often reappear in broadly similar positions; Fig.~\ref{fig:30mA}(e). We again applied a current to the Pt heater, but this time with a negative current $I = -30$~mA. Fig.~\ref{fig:30mA}(f), shows that a negative current also drives the domains towards the heater, this time by several tens of nm over two minutes. 
 
To quantify the domain motion, we generated line profiles through the domains by integrating 50 pixel wide strips (corresponding to 570~nm). Examples are shown in Fig.~\ref{fig:30mA}(g). The dips in the line profiles correspond to the contrast from both domains that are mobile (labeled `D') and the previously identified defect positions (labeled `d'). We aligned the images using the defect positions and fitted  Gaussians to the most prominent dips, shown with solid lines. The centres of the fitted Gaussians are taken as the positions of the different features. As expected, we do not see motion outside the fitting uncertainty for the defects. We take the change in position of a domain being its displacement under the effect of $\nabla T$ during the 120~s that the heater current flowed. Taking the domain motion between the black and orange line profiles as an example, which moved $-120 \pm 30$~nm, this yields an average velocity of $v \approx -0.9 \pm 0.2$~nm/s. (Motion towards the heater is in the negative $x$ direction.) A table giving all of the fitting parameters for the line profiles shown in Fig.~\ref{fig:30mA}(g) is presented in the Supplementary Material \cite{suppmater}.

We find that the domains always move towards the hotter region, but their average velocity whilst the heater current $I$ flows depends its direction, as in the example shown in Fig.~\ref{fig:30mA}. An explanation for this difference is that it is due to the Oersted field generated by $I$. This field will be vertical where it passes through the magnetic track and will become weaker with distance from the Pt heater according to the Biot-Savart law. The sign of $I$ determines the sign of this field and its gradient, which can enhance or diminish any effect on the motion of the domain due to $\nabla T$. We discuss this point in more detail in the Supplementary Material~\cite{suppmater}. In our experiments the domain wall always moves towards the heater, so the net force in this direction must always be larger here than any opposing effect from the Oersted field. The displacement is indeed larger for $I<0$ and smaller for $I>0$ for positive $H$, consistent with our observations. 

\begin{figure}[t]
    \includegraphics[width=7cm]{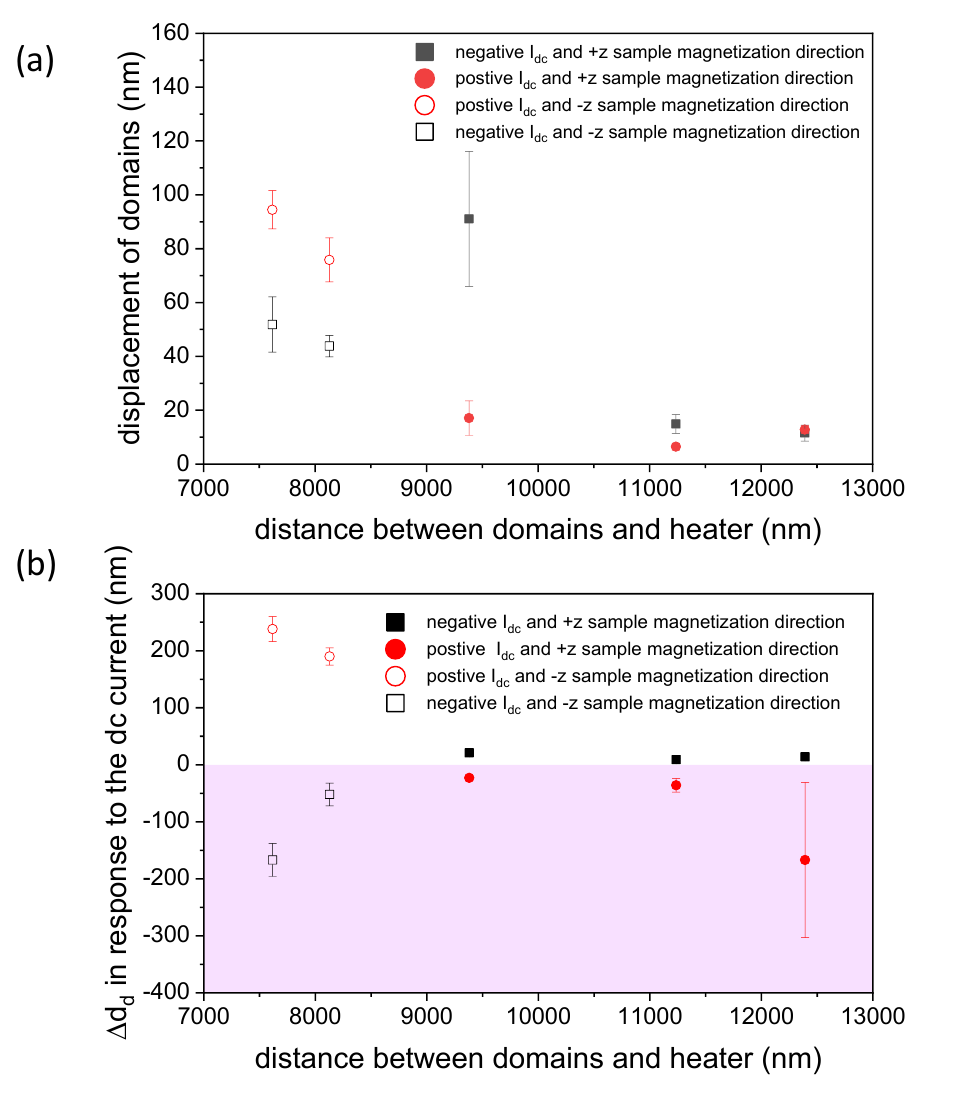}
    \caption{Effects of changing sign of $I$ and $H$. (a) Domain motion and (b) expansion/contraction. The domains were driven by heater currents $I = \pm 30$~mA current for both magnetization directions stabilized in fields $H = \pm 30$~Oe.}
    \label{fig:pndomain}
\end{figure}

We confirmed the effect of the Oersted field from the heater wire by inverting the magnetization in the domains by re-nucleation with opposite field $H$. This will cause the Oersted field forces to invert. In Fig.~\ref{fig:pndomain}(a), we show data for a variety of domains at different distances from the heater and for different combinations of $I$ and $H$. As previously, $I = \pm 30$~mA, $H = \pm 30$~Oe, and the current was applied for two minutes. Although the domains always move towards the heater, they consistently move further when the Oersted field gradient gives rise to an attractive force ($H$ and $I$ have opposite signs) and less far when the force is repulsive ($H$ and $I$ have the same sign). 

Whilst so far we have implicitly described the domains as rigid objects, they are of course regions of reversed magnetization between two domain walls, and it is in fact these walls upon which temperature gradients will act. We can see from the data in Fig.~\ref{fig:pndomain}(a) that the domains move at significantly higher average speeds if they are closer to the heater, implying that the temperature gradient is steeper there. This means that there will be a small difference in the magnitude of $\nabla T$ for the walls on either side of each domain. Given the $1/x$ dependence of the Oersted field, its gradient will also be slightly different on each of the two walls. By examining the widths of the fitted Gaussians we can hence study any domain expansion or contraction that arises from differential domain wall movement at either side of the domain. The results are shown in Fig.~\ref{fig:pndomain}(b), where $\Delta d_\mathrm{d}$ is the change in domain size. The data show that the domains expand when $I$ and $H$ have opposite sign, and contract when they have the same sign. Again, the effects are stronger when the domains are closer to the heater, where the gradients are stronger. These results are consistent with the force model shown in Fig.~S3 in the Supplementary Material \cite{suppmater} and thus further support this picture. 

\begin{figure}[t]
    \includegraphics[width=7cm]{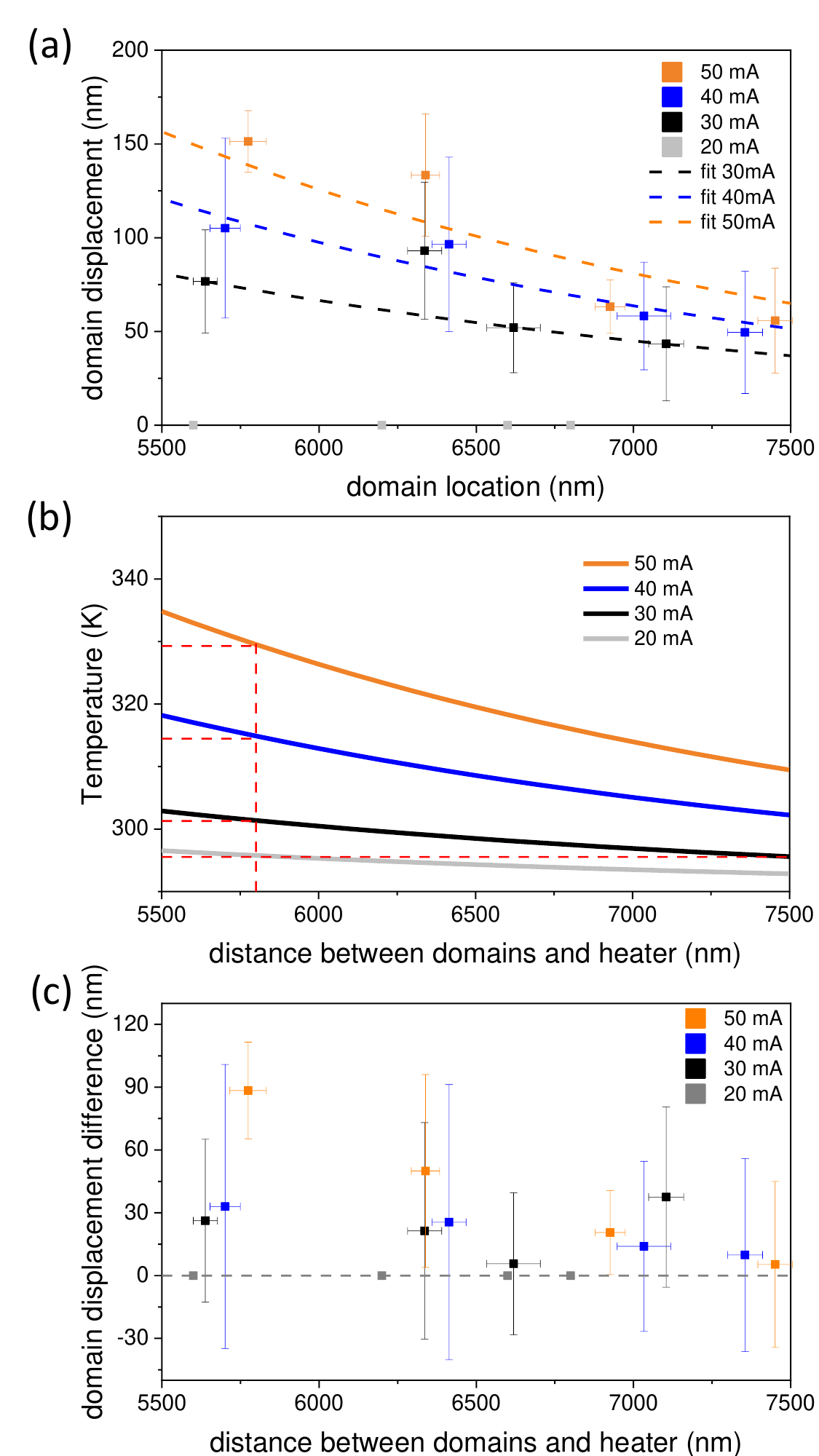}
    \caption{Temperature gradient-driven domain motion. (a) Average of the domain displacement after cancellation of Oersted field effects. (b) Estimated temperature profiles along the magnetic track. The red dashed lines point out the different temperatures at $x= 5800$~nm, the position of a defect, generated by different dc currents. (c) The difference of the displacement of the domain driven by varying positive and negative currents.}
    \label{fig:domaintemp}
\end{figure}

In order to extract the motion arising only from $\nabla T$, we average the measurements for a given re-nucleated domain for both positive and negative heater current, which will cancel out the Oersted field gradient-induced contribution to the motion. The resulting relationship between the $\nabla T$-induced domain displacement and distance from the heater is shown in Fig.~\ref{fig:domaintemp}(a) for a range of heater currents between 20 and 50~mA. The abscissa shows the distance from the domain to the heater, and the error bar is the standard deviation of the positions of four different domains after every pulse (three repeat pulses for each condition). The ordinate shows the domain displacement, and the error bar is the standard deviation of the displacement of domains after every pulse. There is considerable variation in the data caused by sample non-uniformities. Nevertheless, the main trends are clear. For $I = 20$~mA we observed no displacement to within the uncertainty of our measurement. Motion sets in when the current is 30~mA or higher, and larger displacements are observed when the current rises further. 

As in Fig.~\ref{fig:pndomain}, another clear trend in Fig.~\ref{fig:domaintemp} is that larger displacements (faster motion) is observed for domains close to the heater. This implies that the temperature gradient is steeper there. A simple assumption is that temperature varies exponentially with position $x$ along the track as heat conducts away into the substrate, which we can quantify as
\begin{equation}
    T(x) = T_\mathrm{lab} + \Delta T \exp \left(-\frac{x}{l_\mathrm{T}}\right), \label{eq:tempprofile}
\end{equation}
where $x = 0$ is the edge of the heater wire, $T_\mathrm{lab} = 290$~K is the laboratory ambient temperature, $\Delta T$ is the temperature rise for the heater wire taken from the Pt heater calibration (see Supplementary Material \cite{suppmater}), and $l_\mathrm{T}$ is the characteristic lengthscale for a $1/e$ decay of the temperature. Differentiating yields $\nabla T = -(\Delta T / l_\mathrm{T}) \exp (-x/l_\mathrm{T})$. We then define the thermal efficiency $k$ in terms of the average domain displacement $\Delta x_\mathrm{d} = k \nabla T$, since DW velocity is expected to be proportional to $\nabla T$ \cite{Chico2014}. We fitted this expression to our data as shown in Fig.~\ref{fig:domaintemp}(a) to obtain $k = 4700 \pm 600$~nm$^2$K$^{-1}$ and $l_\mathrm{T} = 2400 \pm 200$~nm. These values were then used to construct the $T(x)$ profiles shown in Fig.~\ref{fig:domaintemp}(b) from Eq.~\ref{eq:tempprofile}.

For our highest $I = 50$~mA, the temperature at the position of a prominent defect---$x = 5800$~nm, marked with a red dashed line---was lower than 330~K. Thus, we can say that the region imaged here by MFM remained well below the Curie temperature since we can still observe the magnetic pattern, and structural changes to the multilayer are also unlikely for these modest temperature rises. Moreover, $\nabla T = -2.4$~K/$\upmu$m  at $x = 5800$~nm generated by 20~mA (no motion) is just below the $\nabla T = -2.6$~K/$\upmu$m generated by 30~mA current at $x = 7200$~nm (slowest motion observed). The threshold temperture gradient for the onset of motion in our system therefore lies between these two values, $\nabla T_\mathrm{threshold} = -2.5 \pm 0.1$~K/$\upmu$m.

In Fig.~\ref{fig:pndomain} (c), we plot the difference in domain displacement for positive and negative $I$, which is a measure of the strength of the Oersted field gradient, which we might expect to go as $\partial H/\partial x \sim -x^{-2}$ by differentiating the Biot-Savart law. Whilst the error bars on the data are too large to fit the data to this form precisely, the expected broad trends that the difference in displacements is larger when closer to the current-carrying wire, and when the $I$ is larger, can be seen in the data.

The order of magnitude of the effect we find is an average velocity $\sim -1$~nm/s in a temperature gradient of $\sim -20$~K/$\upmu$m -- very slow and hence expected to be in the creep regime \cite{Metaxas2007}. (Negative velocity indicates motion towards the heater, the edge of which is at $x = 0$.) We estimate the strengths of different mechanisms that can contribute to this $\nabla T$-driven domain motion: more details of the following calculations are given in the Supplementary Material \cite{suppmater}. 

For the spin Seebeck effect, we follow the method of Jiang \textit{et al.} \cite{Jiang2013}, who write the velocity of a domain wall under a number current density $J_\mathrm{m}$ of magnons as
\begin{equation}\label{vmag}
  v = -\frac{\beta}{\alpha} \frac{\gamma \hbar}{M_\mathrm{s}} J_\mathrm{m},
\end{equation}
where $\gamma = 1.76 \times 10^{11}$~Hz/T is the gyromagnetic ratio, $M_\mathrm{s}$ is the saturation magnetization, $\beta$ is the nonadiabaticity parameter and $\alpha$ the Gilbert damping. For realistic materials parameters~\cite{suppmater} we estimate $J_\mathrm{m} = 1.8 \times 10^{27}$~m$^{-2}$s$^{-1}$. Putting this value into Eq.~\ref{vmag} leads to $v = -6.2$~cm/s. 

Meanwhile for the spin-dependent Seebeck effect, for a pure electronic spin current density $J_\mathrm{s}$, we can write~\cite{suppmater}
\begin{equation}\label{velec}
  v = -\frac{\beta}{\alpha} \frac{\gamma \hbar}{2 e M_\mathrm{s}} J_\mathrm{s}.
\end{equation}
Determining $J_\mathrm{s}$ using the expression given by Yi \textit{et al.} \cite{Yi2020}, we obtain $v = -27$~cm/s.

The entropic contribution comes from the temperature, and therefore spatial, variation of the magnetic parameters and can be derived from the LLB equation~\cite{Schlickeiser2014}, giving
\begin{equation}
    v = -\frac{2\gamma}{\alpha M_0}\left( 1 + \alpha^2\left(\frac{M_s}{M_0}\right)^2 \right)\left(\frac{\partial A}{ \partial T} \frac{\partial T}{\partial x} \right),
\end{equation}
where $A$ is the exchange stiffness which is a function of temperature. By calculating the $A(T)$ from the temperature dependence of the spin-wave stiffness so that proper Planck statistics are used, this gives a velocity of $v = -1.1$~m/s. 

All three of these velocities are directed towards the hot end of the temperature gradient, which matches our experimental observation. That arising from the entropic force is much the largest, and hence can be expected to be the predominant effect giving rise to the motion that we have observed. 


To summarise, the displacement of reverse domains in a perpendicularly magnetized metallic multilayer driven by a temperature gradient was observed by MFM. The domains always move towards the hotter region. By cancelling the effect of Oersted fields generated by the calibrated thermometer/heater wire we are able to quantify the effect: we see velocities of $\sim 1$~nm/s under a temperature gradient of$\sim 20$~K/$\upmu$m, slow enough to be in the creep regime. By estimating the size of different possible driving mechanisms based on the experimentally measured properties of our multilayer we find that the entropic forces \cite{Schlickeiser2014} are expected to the the predominant effect, confirming LLB-based simulations \cite{Moretti2017}. Cleverly engineering the temperature dependence of the DW energy, e.g. using layers with phase transitions at controlled temperatures or novel scaling relations between micromagnetic parameters, could allow this currently quite weak effect to become large enough to be useful, perhaps operating using scavenged waste heat. 

\begin{acknowledgments}
We wish to thank Philippa Shepley, Nathan Satchell, and Md Golam Hafiz for their experimental assistance. 
\end{acknowledgments}

\bibliography{thermaldomain}

\end{document}


\title{Temperature gradient-driven motion of magnetic domains in a magnetic metal multilayer: Supplementary material}
\author{Lin~Huang}
\altaffiliation[Present address: ]{School of Material Science, University of Sheffield, Sheffield S10 2TN, United Kingdom.}
\affiliation{School of Physics and Astronomy, University of Leeds, Leeds LS2 9JT, United Kingdom.}

\author{Joseph Barker}
\affiliation{School of Physics and Astronomy, University of Leeds, Leeds LS2 9JT, United Kingdom.}
  
\author{Lekshmi~Kailas}
\affiliation{School of Physics and Astronomy, University of Leeds, Leeds LS2 9JT, United Kingdom.}

\author{Soumyarup Hait}
\affiliation{School of Physics and Astronomy, University of Leeds, Leeds LS2 9JT, United Kingdom.}

\author{Simon~D.~Connell}
\affiliation{School of Physics and Astronomy, University of Leeds, Leeds LS2 9JT, United Kingdom.}

\author{Gavin~Burnell}
\affiliation{School of Physics and Astronomy, University of Leeds, Leeds LS2 9JT, United Kingdom.}

\author{Christopher~H.~Marrows}
\email[Email:~]{c.h.marrows@leeds.ac.uk}
\affiliation{School of Physics and Astronomy, University of Leeds, Leeds LS2 9JT, United Kingdom.}

 


\maketitle

\section{Sample patterning, calibration of platinum heater/thermometer, and imaging}

The current contacts were Ti~(5~nm)/Au~(100~nm) bilayers, deposited by thermal evaporation, which were patterned by photolithography. The bilayer photoresists were S1813 (spin coated at 300~rpm and baked for 180~s on a 185$^\circ$C hot plate) and LOR3B (spin coated at 3000~rpm and baked for 60~s on a 115$^\circ$C hot plate).

\begin{figure}[b!]
    \centering
    \includegraphics[width=10cm]{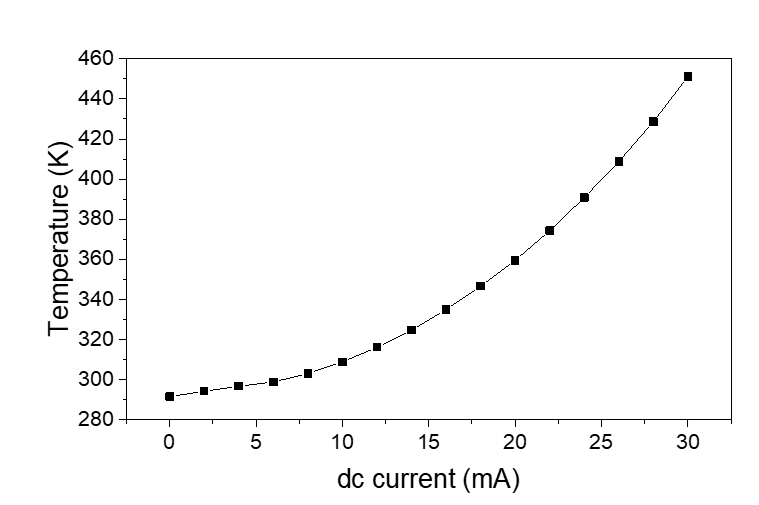}
    \caption{Temperature reached by Pt heater as a function of heater current.}
    \label{fig:PtT}
\end{figure}

To generate the temperature gradient along the magnetic track, a dc current was passed through the Pt heater in order to heat the magnetic track at the point where the track and heater cross. The dc current for heater was injected with a Keithley 6221 current source. To perform the calibration of the Pt thermometer/heater resistance, the ac resistance of the Pt heater was measured using a SR830 lock-in amplifier, with the ac sense current also supplied by the Keithley 6221. The `K' shape of the Pt heater means that we could make a four-point resistance measurement for thermometry at that point. We measured the temperature coefficient of resistance of the Pt wire and then used the measured steady-state resistance rise for a given dc current to generate the calibration curve shown in Fig.~S\ref{fig:PtT}. The laboratory temperature, and hence the ambient temperature of the magnetic track far from the heater, was around 290~K.

The samples were imaged by atomic (AFM) and magnetic force microscopy (MFM) using an Asylum Research MFP-3D microscope, using ASYLMFLC-R2 magnetic coated MFM tips. Out-of-plane magnetic fields were applied by means of a Variable Field Module VFM3, which uses permanent magnets and so avoids any inadvertent sample heating. 

\section{Sheet film hysteresis curve}

\begin{figure*}[t]
    \centering
    \includegraphics[width=10cm]{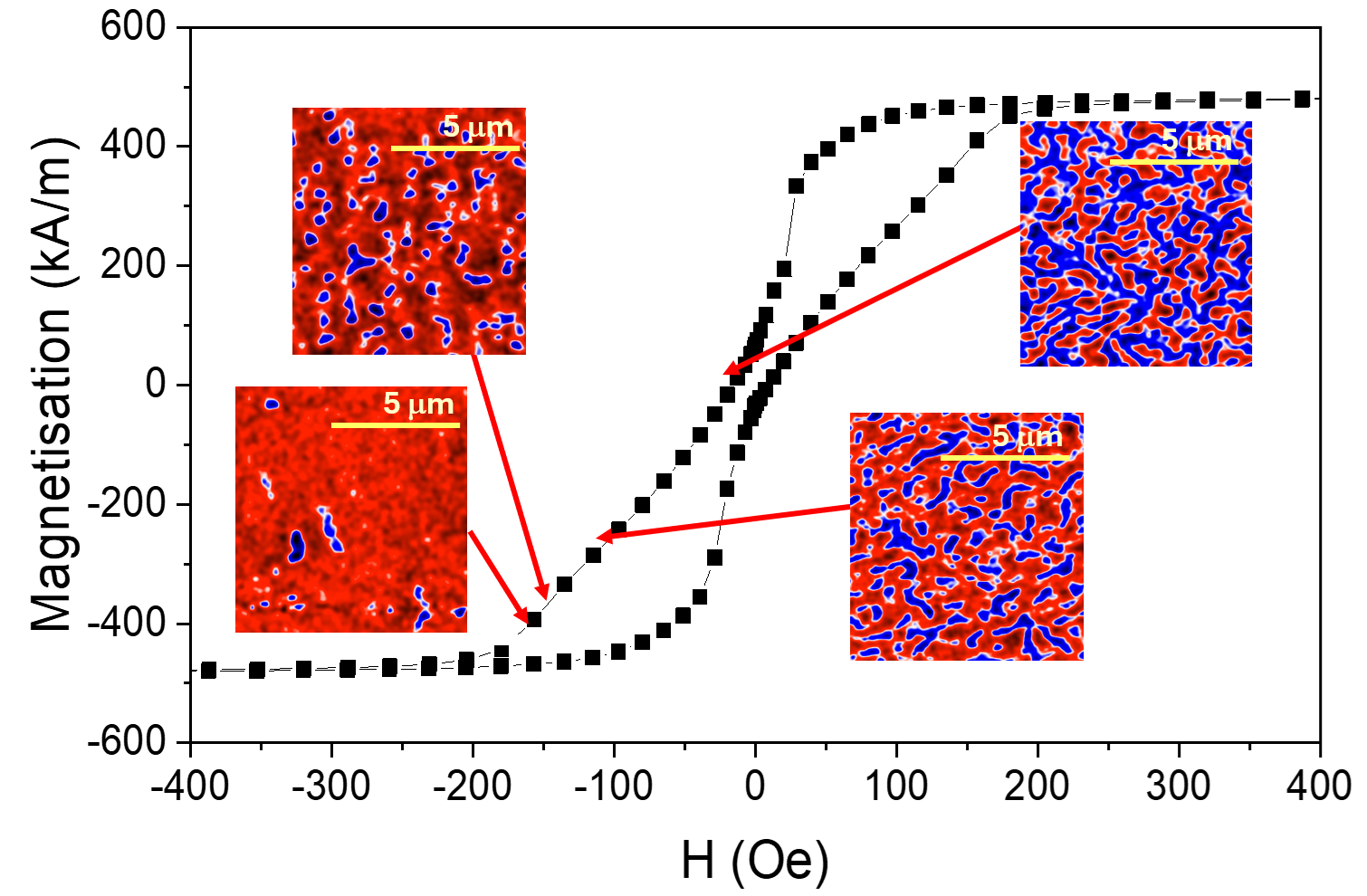}
    \caption{Hysteresis loop under out-of-plane field for a sheet film Ta(20~\AA)/[Pt(7~\AA)/Co$_{68}$B$_{32}$(8~\AA)/Ir(5~\AA)]$_{\times 10}$/Pt(14~\AA) multilayer subjected to the same heating processes that occur during track fabrication. False colour MFM images show the magnetic domain patterns for selected positions on the loop, indicated by the red arrows.}
    \label{fig:mfms}
\end{figure*} 

In Fig.~S\ref{fig:mfms}, we show an out-of-plane hysteresis loop measured by SQUID-vibrating sample magnetometry for a sheet film of multilayer subjected to the same heating processes that occur during track fabrication. Its low saturation field shows that there is preferentially perpendicular magnetisation that reverses through domain formation processes. Some of the associated domain structures at selected values of field, observed by magnetic force microscopy (MFM), are shown as insets. 

At the coercive field, which is also close to remanence, the sample is equally divided into up and down domains with an interconnected labyrinthine appearance. As the reverse field is made stronger, the forward-magnetized domains become discrete objects, some with a circular form, others with an extended worm-like appearance. These become fewer and fewer in number until approaching the saturation magnetic field where only a few remain. Even at this point, where the temperature gradient-driven motion experiments were carried out, many domains still have an extended appearance. 

\section{MFM image processing}

\begin{figure*}[t]
    \centering
    \includegraphics[width=16cm]{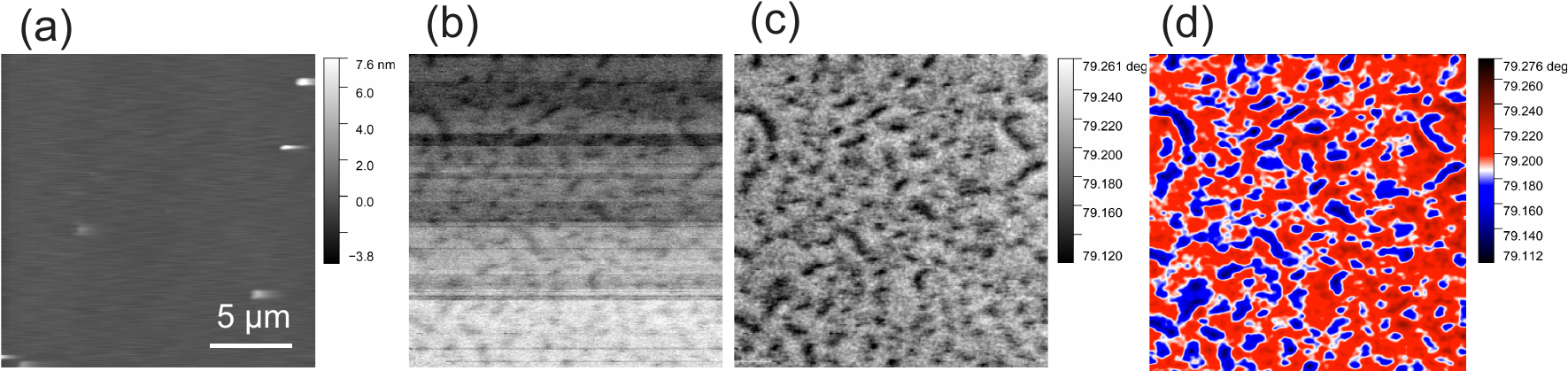}
    \caption{An example of the MFM image processing workflow, for an image acquired at -135~Oe. This is an additional image from the hysteresis loop shown in Fig.~S\ref{fig:mfms}. (a) Height image, (b) raw MFM image (`NapPhaseRetrace') acquired on Asylum Research MFP-3D equipped with a Variable Field Module VFM3. (c) Gray scale MFM image line-levelled to $0^\mathrm{th}$ order (no masking) and contrast enhanced. (d) Final image is smoothed with a 1 pixel width Gaussian filter in Gwyddion, and a blue-red color palette applied.}
    \label{fig:MFM_workflow}
\end{figure*} 

In Fig.~S\ref{fig:MFM_workflow} we show the workflow used for MFM image processing, which was carried out in part by using the Gwyddion software package \cite{Necas2012}. The process involved, line-levelling, contrast enhancement, and smoothing, before a blue-red color palette is applied. 

\section{Gaussian fitting}

The positions and widths of the Gaussians fitted to the dips in contrast in the line profiles shown in Fig.~1(g) of the main text, before and after the application of the $-30$~mA current, are given in Table~\ref{tab:peak30mA}. Contrast dips are either labelled as defects `d' when occupying positions identified in the MFM image under a saturating field in Fig.~1(b), or domains `D' when motion is detected. 

\begin{table}[b]
\caption{\label{tab:peak30mA}
Gaussian fitting results for the contrast dips in the line profiles along the magnetic track shown in Fig.~1(g). These correspond to before (profiles 1 and 3) and after (profiles 2 and 4) application of a -30~mA heater current $I$ for two minutes at two different positions across the track width}

\begin{ruledtabular}
\begin{tabular}{ccccc}

Line Profile (colour)&Feature & Application & Gaussian & Gaussian \\
 & &  of current &  width (nm) & centre (nm) \\
\hline
1 (Black)&defect d1 & before& $390 \pm 30$ & $330 \pm 10$ \\
2 (Orange)&defect d1 & after & $390 \pm 20$ & $330 \pm 10$ \\
\hline
1 (Black)&domain D1 &before & $130 \pm 10$ & $1840 \pm 10$ \\
2 (Orange)&domain D1 & after& $200 \pm 20$ & $1720 \pm 30$ \\
\hline
1 (Black)&defect d2 & before & $180 \pm 20$ & $2370 \pm 10$ \\
2 (Orange)&defect d2 & after & $160 \pm 10$ & $2370 \pm 10$ \\
\hline
3 (Blue)& defect d3 + domain D2 & before & $ 400 \pm 10 $ & $ 310\pm 20 $ \\
4 (Pink)&defect d3 + domain D2 & after & $270 \pm 20 $ & $230\pm 10 $ \\
\hline
3 (Blue)&domain D3&before & $ 150\pm 10 $ & $ 1270\pm 10 $ \\
4 (Pink)&domain D3 & after& $150\pm 40  $ & $1200 \pm 10 $ \\
\hline
3 (Blue)&defect d4 + domain D4  & before & $ 160\pm20 $ & $ 2420\pm 10 $ \\
4 (Pink)&defect d4 + domain D4 & after & $220\pm 20$ & $2390\pm 10$ \\
\hline
3 (Blue)& defect d5 & before& $ 200\pm 20   $ & $ 3070\pm20 $ \\
4 (Pink)& defect d5 & after & $ 350\pm 20 $ & $ 3090\pm 40 $ \\
\end{tabular}
\end{ruledtabular}
\end{table}

For the contrast dips that are pure defects (d1, d2, and d5), no motion of the fitted Gaussian centre is detected within the uncertainty of the fits. Domains D1 and D3 are found away from the defect positions and the motion of the fitted gaussian centre is clearly the motion of the domain itself. In some cases (d3+D2 and d4+D4) we have a contrast dip that is at a position where a defect was identified but we also see motion of the fitted gaussian centre. We take these to be domains that have (not unexpectedly) nucleated at defects and have been subjected to driving forces arising from the temperature gradient that have caused their positions to shift. 

\section{Effects of heater-generated Oersted fields}

 A current in the heater wire will give rise to both heating and an Oersted field, both of which decay with distance along the track from the heater wire position. We consider how the Oersted field gradient will affect magnetic textures in the track in combination with the temperature gradient. 
 
\begin{figure*}[t]
    \centering
    \includegraphics[width=14cm]{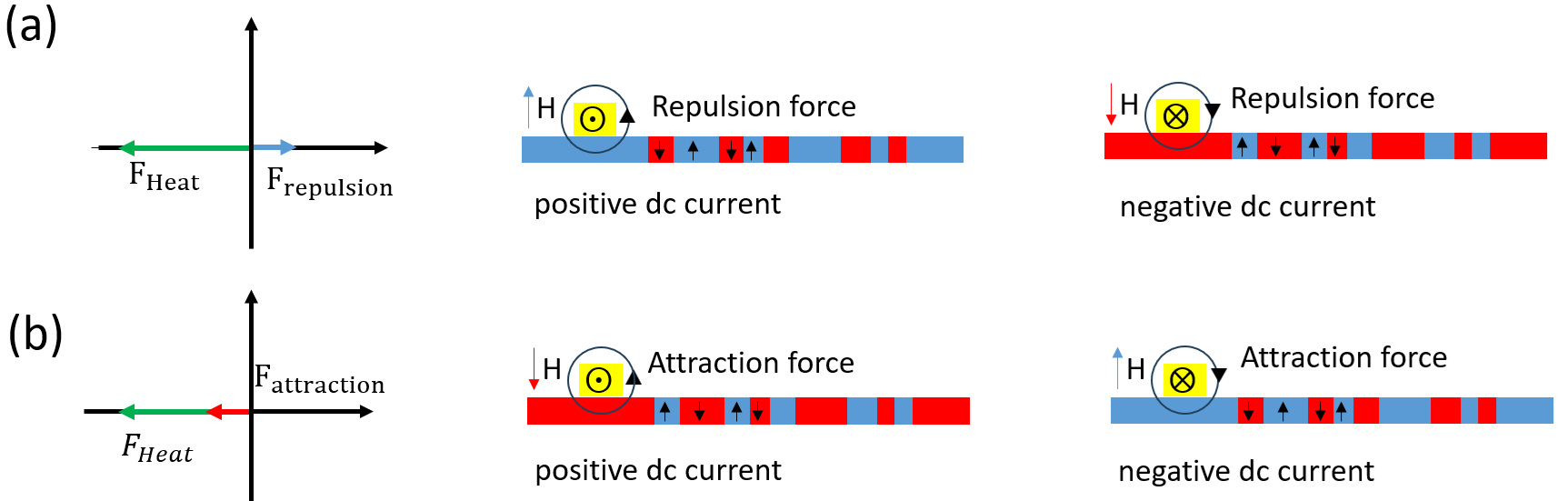}
    \caption{Forces on reverse domains considered in a side view of the track. Heater currents flow into or out of the plane of the paper. (a) A repulsive force on reverse domains can be generated by the Oersted field gradient when the Oersted field direction in the track matches the externally applied field ($H$) direction. (b) An attractive force on reverse domains can be generated by the Oersted field gradient when the Oersted field direction in the track opposes the externally applied field ($H$) direction.}
    \label{fig:oersted}
\end{figure*}

 Taking the first case shown in Fig.~S\ref{fig:oersted} (a), the wire is preferentially magnetized by a vertical upwards externally applied field $H$ (blue) but still contains some reverse domains (red). For a positive dc current $I$ in the heater (yellow) the moment in these domains opposes the circulating Oersted field, and so the seeks a weaker Oersted field, which is further from the wire. Hence the Oersted field gradient generates a repulsive force on the domain $F_\mathrm{repulsion}$, which opposes the thermal force $F_\mathrm{Heat}$ if we assume that that force causes the domain to move toward the hot end of the gradient. 
 
 Reversing both the sense of both $H$ and $I$ will still lead to a repulsive force from the Oersted field gradient. Reversing only one or the other will give rise to an attractive force Oersted field force $F_\mathrm{attraction}$, Fig.~S\ref{fig:oersted}(b), which enhances the force from the temperature gradient $F_\mathrm{Heat}$.

 Hence, this picture is consistent with our experimental observation that the domain motion is always towards the heater, but varies depending on the relative alignment of the heater current direction and magnetisation within the domains. 

\section{Quantitative comparison of domain motion driving mechanisms
\label{sec:drivingmechanisms}}

In the experiments reported in this Letter, domain velocities of $\sim 1$~nm/s under a thermal gradient of 20~K/$\upmu$m were observed in perpendicularly magnetised Pt/CoB/Ir multilayers. The motion was towards the hot end of the gradient, i.e. the domain velocity $v$ has the same sign as the temperature gradient $\nabla T$. Here we estimate the relative size and directions of different possible driving mechanisms in an attempt to determine which is most likely to be dominant in situations like this one. We consider (i) motion driven by the spin-transfer torque (STT) exerted by a current of magnons arising from the spin Seebeck effect; (ii) motion driven by STT arising from the pure spin current of electrons caused by the spin-dependent Seebeck effect in metals, and (iii) entropic forces arising from the gradient in free energy of the spin textures. 

Most theories of domain wall motion deal with the steady-state velocity of the wall in the viscous flow regime. Since this velocity is proportional to the driving force on the domain wall \cite{Thiele1973,Thiaville2005,Marrows2005} we can use the relative sizes of these velocities, when calculated for the various mechanisms of interest, as measures of the relative sizes of the driving forces on the domains in our experiment. 

\subsection{Magnon currents}

For the spin Seebeck effect, we follow the method of Jiang \textit{et al.} \cite{Jiang2013}, who write the velocity of a domain wall under a current density $J_\mathrm{m}$ of magnons as
\begin{equation}\label{vmag}
  v = -\frac{\beta}{\alpha} \frac{\gamma \hbar}{M_\mathrm{s}} J_\mathrm{m},
\end{equation}
where $\gamma = 1.76 \times 10^{11}$~Hz/T is the gyromagnetic ratio. A typical value for the Gilbert damping in such a multilayer is $\alpha \approx 0.07$ \cite{Zeissler2020}, and since the nonadiabaticity parameter $\beta$ is essentially unknown but typically of the same order of magnitude, we set $\beta = \alpha$ since they are likely to be of the same order of magnitude. We use a value of $M_\mathrm{s} = 532$~kA/m for the saturation magnetisation of CoB at $\sim 325$~K based on our SQUID-VSM measurements.

The formula given in Ref.~\onlinecite{Jiang2013} for the magnon current density is
\begin{equation}\label{magcurrent}
  J_\mathrm{m} = - \frac{k_\mathrm{B} \nabla T}{6 \pi^2 \lambda \hbar \alpha} F_0,
\end{equation}
in which ``$F_0 \approx 1$ is a coefficient arising in the kinetic theory of magnonic STT \cite{Kovalev2012,Kovalev2009}''. The remaining material parameter to be determined is the thermal magnon wavelength $\lambda$. We estimate this using \cite{Cornelissen2016}
\begin{equation}\label{lambdathermal}
  \lambda = \sqrt{\frac{D_0}{k_\mathrm{B} T}},
\end{equation}
where $D_0 = 5.65 \times 10^{-40}$~J/m$^2$ is the spin wave exchange stiffness, determined from a fit to the SQUID-VSM data for our film (see \ref{sec:fit_stiffness}). This value yields $\lambda = 0.36$~nm at $T=325$~K.

Inserting this value for $\lambda$ into Eq.~S\ref{magcurrent} yields $J_\mathrm{m} = 1.8 \times 10^{27}$~m$^{-2}$s$^{-1}$ for $\nabla T = -20$~K/$\upmu$m. (Typical electron current densities for STT are $\sim 10^{12}$~A/m$^2$, which corresponds to a particle flux density of $\sim 6 \times 10^{30}$~m$^{-2}$s$^{-1}$.)

Putting this value for $J_\mathrm{m}$ into Eq.~S\ref{vmag} leads to \framebox[1.1\width]{$v = -6.2$~cm/s}. This value agrees in sign with the experimental observation but is many orders of magnitude larger. This is perhaps not surprising since the expressions above assume a viscous flow regime, whereas the experiment is in the creep regime. Nevertheless, this value can act as a measure of the strength of the STT arising from the magnon current. 

\subsection{Electronic pure spin currents}

For a conventional electrical charge current density $J_\mathrm{c}$, the velocity of a DW driven by STT is usually written as~\cite{Marrows2005,Thiaville2012,Jiang2013} 
\begin{equation}
    v = -\frac{\beta}{\alpha} \frac{\gamma \hbar P}{ 2 e M_\mathrm{s}} J_\mathrm{c},
\end{equation}
where $P = (\sigma_\uparrow - \sigma_\downarrow)/(\sigma_\uparrow + \sigma_\downarrow)$ is the spin polarization of the current; typically $P \sim 0.5$ in the diffusive regime in Co \cite{Bass1999}. The minus sign ensures that the domain wall velocity is in the direction of electron flow \cite{Thiaville2005}. Since the spin current $J_\mathrm{s} = P J_\mathrm{e}$, we can re-write this as
\begin{equation}\label{velec}
  v = -\frac{\beta}{\alpha} \frac{\gamma \hbar}{2 e M_\mathrm{s}} J_\mathrm{s}.
\end{equation}
The spin-dependent Seebeck effect will drive a pure spin current with equal numbers of spin-$\uparrow$ and spin-$\downarrow$ electrons moving in opposite directions, that is to say that $J_\uparrow = - J_\downarrow$, ensuring that the net charge current $J_{e} = J_\uparrow + J_\downarrow = 0$, as required by the open circuit configuration of our experiment. Meanwhile, $J_\mathrm{s} = J_\uparrow - J_\downarrow$.

To determine the electronic spin current we use the expression given by Yi \textit{et al.} \cite{Yi2020}
\begin{equation}\label{spincurrent}
  J_\mathrm{s} = - \frac{2 \sigma_\uparrow \sigma_\downarrow}{\sigma_\uparrow + \sigma_\downarrow} \left(S_\uparrow - S_\downarrow  \right) \nabla T,
\end{equation}
where $\sigma_{\uparrow,\downarrow}$ are the spin-resolved electrical conductivities and $S_{\uparrow,\downarrow}$ are the spin-resolved Seebeck coefficients. We can obtain the spin-resolved conductivities from knowledge of the resistivity of a material (here we use the value $\rho = 21~\upmu \Omega$cm measured for our multilayer as a whole and assume that the magnetic layers have this stack-averaged resistivity) combined with a value for $P$. Using $P = 0.5$, we obtain
\begin{equation}
  \sigma_\uparrow = \frac{1+P}{2 \rho} = 3.6 \times 10^6 \Omega^{-1}\mathrm{m}^{-1} \quad \mathrm{and} \quad\sigma_\downarrow = \frac{1-P}{2 \rho} = 1.2 \times 10^6 \Omega^{-1}\mathrm{m}^{-1}.
\end{equation}

The spin-dependent Seebeck coefficients of CoB are unknown. As an upper limit, we use the values for Co obtained in the experiments of Yang \textit{et al.} \cite{Yang2021}, $S_\uparrow = -2~\upmu$V/K and $S_\downarrow = -73~\upmu$V/K. Inserting all these values in Eq.~S\ref{spincurrent}, for a gradient of $-20$~K/$\upmu$m, we obtain a pure spin current of $J_\mathrm{s} = 2.5$~GA/m$^2$. 

Taking the same values for $\alpha$, $\beta$, and $M_\mathrm{s}$ as for the spin Seebeck magnon currents, we obtain, from Eq.~\ref{velec}, \framebox[1.1\width]{$v = -27$~cm/s}. Again the sign of the effect matches the observed motion, but the effect is a few times stronger than that arising from the magnon current. We still have the same discrepancy in magnitude between the creep and viscous flow regimes. 

\subsection{Energy gradients}

The domain wall velocity due to gradients in the micromagnetic free energy parameters can be derived from the Landau-Lifshitz-Bloch equation~\cite{Schlickeiser2014}. Below the Walker breakdown, the velocity is
\begin{equation}
    v = \frac{\gamma}{\alpha M_0}\left(1 + \alpha^2\left(\frac{M_0}{M_s}\right)^2 \right)\left( \frac{M_s  B_z \Delta}{\pi} - 2\frac{\partial A}{\partial x} \right).
\end{equation}
In our experiments the domains do not move with an applied field, so we neglect the field term in this equation, leaving
\begin{equation}
    v = -\frac{2\gamma}{\alpha M_0}\left(1 + \alpha^2\left(\frac{M_0}{M_s}\right)^2 \right)\frac{\partial A}{\partial x}.
    \label{eq:entropy_velocity}
\end{equation}
The derivative of the spatial dependence of the exchange stiffness can be related to the temperature gradient through the chain rule
\begin{equation}
    \frac{\partial A}{\partial x} = \frac{\partial A}{\partial T}\frac{\partial T}{\partial x}.
\end{equation}
To obtain $\partial A/\partial T$, we need the temperature dependence of $A$. This is often written as a power law in the reduced magnetisation $A(T) = A_0 \left(M_s/M_0 \right)^\alpha$ where $\alpha=2$ in the mean field picture~\cite{Atxitia_PhysRevB_82_134440_2010}. However, this result is derived for classical Rayleigh-Jeans statistics rather than Planck statistics that are already assumed within Bloch's law and therefore $M_s$. To have a consistent approach, we therefore calculate the temperature dependence of $A$ from the temperature dependence of the spin wave stiffness~\cite{kittel_quantum_theory_solids},
\begin{equation}
    D(T) = D_0\left(1-\frac{U_T}{2U_0}\right),
\end{equation}
where $U_0 = 3 D_0 M_0 / (\hbar\gamma \delta^2)$ is the energy density per spin, with $\delta$ the nearest neighbour distance between atoms, and $U(T)$ is the energy density of the magnon gas at temperature $T$. The exchange stiffness is related to the spin wave stiffness by
\begin{equation}
    A(T) = \frac{M_s}{2\hbar\gamma}D(T).
\end{equation}
By the product rule, we obtain
\begin{equation}
    \frac{\partial A}{\partial T} = \frac{1}{2\hbar\gamma}\left(M_s \frac{\partial D}{\partial T} + \frac{\partial M_s}{\partial T}D \right).
\end{equation}
For the very thin films in these experiments, Bloch's law in the usual `bulk' form cannot be used~\cite{Nembach_NatPhys_11_825_2015,Mohammadi_ACSApplElectronMater_1_2025_2019,Alshammari_PhysRevB_104_224402_2021}. The quantisation of the spin waves in the thin direction leads to significant differences in the equations and the value of stiffness extracted from experiments. The thin film form of Bloch's law is
\begin{equation}
  	M_s(T) = M_0 + \frac{\gamma \hbar}{4\pi L_z}\frac{k_B T}{D_0}
  	\sum_{m = 0}^{N-1} \ln\left[1-\mathrm{exp}\left(-\frac{\omega_m}{k_B T}\right)\right],
    \label{eq:bloch_thin_film}
\end{equation}
where
\begin{equation}
   \omega_m = \gamma\hbar B_z - D_0\left(\frac{m\pi}{(N-1)a}\right)^2 .
\end{equation}
This expression is cumbersome to take derivatives of, so we approximate the zeroth term of the sum using $\ln(1-\exp(-x)) \approx \ln{x}$ as $x \rightarrow 0^{+}$, and using the fact that $N$ is a small number such that the first term of the summation is the dominant term and use the series expansion $\ln(1-e^{-x}) = -\sum
_{n=1}^
{\infty} 
 e^{-nx}/n$. Our approximate form is
\begin{equation}
  	M_s(T) \approx M_0+\frac{\gamma \hbar}{4\pi L_z}\frac{k_B T}{D_0}
  	 \left[\ln{\frac{\omega_0}{k_B T}} - \mathrm{exp}\left(- \frac{\omega_1}{k_B T}\right)\right].
\end{equation}
The derivative is then
\begin{equation}
\frac{\partial M_s}{\partial T} \approx \frac{\gamma \hbar k_B}{4 \pi L_z  D_0}\left[
\ln\left(\frac{\omega_0}{k_BT} \right)
- \mathrm{exp}\left(- \frac{\omega_1}{k_B T}\right) 
- \frac{\omega_1}{k_B T}\mathrm{exp}\left(- \frac{\omega_1}{k_B T}\right) 
- 1
\right].
\end{equation}
We can similarly derive an expression for the internal energy density of the magnon gas in a thin film, finding
\begin{equation}
      	U(T) = \frac{(k_B T)^2}{4\pi L_zD_0}
  	\sum_{m = 0}^{N-1}
  	\operatorname{Li}_2\left[\exp\left(-\frac{\omega_m}{k_B T}\right) \right] - \frac{\omega_m}{k_B T}
  	\ln\left[1-\exp\left(-\frac{\omega_m}{k_B T}\right)\right],
\end{equation}
where $\operatorname{Li}_2(x)$ is the dilogarithm function.
Again, we look to approximate this and write an expression that contains only $\omega_0$ and $\omega_1$
\begin{equation}
      	U(T) \approx \frac{(k_B T)^2}{4\pi L_zD_0}\left\{ 
        \operatorname{Li}_2\left[\exp\left(-\frac{\omega_0}{k_B T}\right) \right]
        +\operatorname{Li}_2\left[\exp\left(-\frac{\omega_1}{k_B T}\right) \right]
  	 - \frac{\omega_1}{k_B T}\ln\left[1 - \exp\left(-\frac{\omega_1}{k_B T}\right)\right]
     \right\},
\end{equation}
which is a reasonable approximation within the low temperature limits of Bloch's law. The derivative of this expression is
\begin{align}
\begin{split}
    \frac{\partial U}{\partial T} \approx \frac{k_B^2 T}{2 \pi L_z D_0}\bigg\{ 
    &\operatorname{Li_2}\left[\exp{\left(-\frac{\omega_0}{k_B T}\right)} \right]
    - \frac{\omega_0}{2k_B T} \ln\left[1 -  \exp{\left(-\frac{\omega_0}{k_B T}\right)}\right] \\
    &+ \operatorname{Li_2}\left[\exp{\left(-\frac{\omega_1}{k_B T}\right)} \right] 
    - \frac{\omega_1}{k_B T}\ln\left[1 - \exp{\left(-\frac{\omega_1}{k_B T}\right)} \right]
    +\frac{\omega_1^2}{2(k_B T)^2}\frac{1}{ \exp{\left(-\frac{\omega_1}{k_B T}\right)} - 1}
    \bigg\}.
\end{split}
\end{align}
Using the values $M_0 = 803$~kA/m, $D_0=5.65\times10^{-40}$~Jm$^2$ (from our fit to Bloch's law in \ref{sec:fit_stiffness}), $\delta = 2.51$\AA, $B_z = 30$~Oe, $L_z = 0.8$~nm, $N=4$ in Eq.~\eqref{eq:entropy_velocity} we find \framebox[1.1\width]{$v=-112$~cm/s}.

\subsection{Comparison of driving mechanisms}

To summarise, the velocities induced by spin transfer torques arising from either magnon or pure spin electron currents would be expected to be in the tens of cm/s scale under viscous flow on the basis of these calculations. Meanwhile, the velocity induced by the DW energy gradient is on the scale of m/s for the same regime. This velocity is around four times larger than the velocities induced by the spin Seebeck or spin-dependent Seebeck effects, and so is likely to be the dominant contribution to the temperature gradient-driven motion that we observe. This remains the case even when our observed velocity is many orders of magnitude less, owing to the pinning that keeps our system in the creep regime: the relative size of the driving forces is unaffected. 

\section{Fitting Bloch's Law
\label{sec:fit_stiffness}}

We fitted Eq.~\eqref{eq:bloch_thin_film} to the experimental data from the lowest temperature up to $T=200$~K. For $B_z$ we used the in-plane saturation field from the hysteresis loop at each temperature as a measure of the coercive field of the film. i.e. $B_z \rightarrow B_z(T)$. We fit using the DREAM Markov chain Monte Carlo algorithm as implemented in the \textsc{bumps} package~\cite{bumps}. The settings are burn $=1000$, samples $= 10^6$, init = eps. The full fitting script and results are available in the data upload. The best fit gives $D_0=(5.65^{+1.25}_{-0.55})\times10^{-40}$~Jm$^2$ and $M_0 = (803.33^{+7.27}_{-11.33})$~kA/m where the errors represent the 68\% confidence interval.

\begin{figure}
    \centering
    \includegraphics[width=10cm]{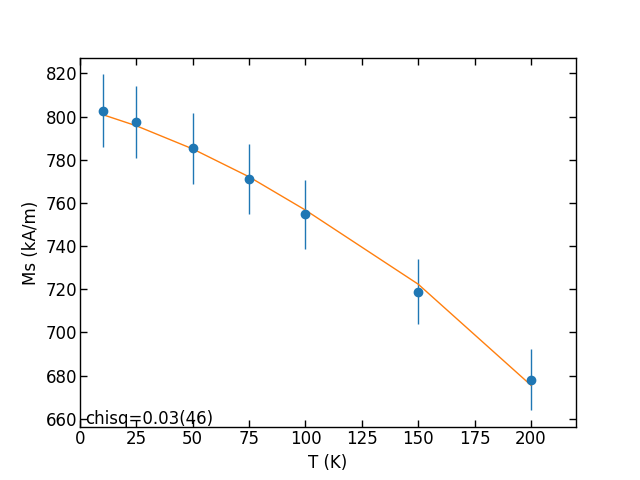}
    \caption{Fitting Eq.~\eqref{eq:bloch_thin_film}, thin film Bloch's law, to the experimentally measured magnetisation. Blue points are the experimental measurements and their error bars. The orange curve is Eq.~\eqref{eq:bloch_thin_film} with the best fit parameters $D_0=(5.65^{+1.25}_{-0.55})\times10^{-40}$~Jm$^2$ and $M_0 = (803.33^{+7.27}_{-11.33})$~kA/m.}
    \label{fig:enter-label}
\end{figure}


\bibliography{thermaldomainsupp}